%% file: qqwh.tex
\documentstyle[aps,12pt,epsf]{revtex} 
\begin{document} 
\pagestyle{plain} 
\setcounter{page}{1} 
\baselineskip=0.3in 
\begin{titlepage} 
\vspace{.5cm} 
 
\begin{center} 
{\Large Top Quark Loop Corrections to the Neutral Higgs \\  
          Boson Production at the Fermilab Tevatron } 

\vspace{.2in} 
Chong Sheng Li \\
 Department of Physics, Peking University,  
Beijing 100871, P.R. China\\ 
\vspace{.4in} 
Shou Hua Zhu  \\
 CCAST (World Laboratory), P.O. Box 8730, Beijing 100080, P.R. China \\
 Institute of Theoretical Physics, Academia Sinica, P.O. Box
 2735, Beijing 100080, P.R. China
\end{center} 

\begin{footnotesize} 
\begin{center}\begin{minipage}{5in} 
\baselineskip=0.25in 
\begin{center} ABSTRACT\end{center}  
We calculate the $O(\alpha m_t^2/m^2_W)$ corrections arising from 
diagrams involving the top-quark loops to the light neutral Higgs boson 
production via $q\bar q'\rightarrow WH$ at the Fermilab Tevatron in both 
the standard model and the minimal supersymmetric model. In contrast 
to the QCD correction which increases the tree-level cross section, 
the corrections 
imply a few percent reduction in the production cross section relative to 
the tree-leve results.   
 
\end{minipage}\end{center} 
\end{footnotesize} 
\vfill 
 
 
\end{titlepage} 
 
\newpage 
\section{Introduction} 
  The Higgs boson is the only particle of the Standard Model(SM) which 
has not been discovered so far. The direct search in the LEP experiments 
via the $e^+e^-\rightarrow Z^*H$ yields a lower bound of $\sim 77.1$ GeV 
on the Higgs mass\cite{a01}. This search is being extended at present LEP2  
experiments, which will explore up to a Higgs boson mass of about $95$ GeV 
via $e^+e^-\rightarrow ZH$ by the year 2000\cite{a02}. Much higher masses will be  
explored by the CERN Large Hadron Collider (LHC). Since it will be some 
years before the LHC comes into operation it is worth considering 
wherther the Higgs boson can be discovered from the existing hadron  
collider, the Tevatron. Much study has been made in the detection 
of a Higgs boson at the Tevatron\cite{a03}.   
It was recently pointed out\cite{a04} that  
a light Higgs boson of mass $60$ GeV $\leq m_H \leq 130 $ GeV can be observable 
at the Tevatron with CM energy $\sqrt{s}=2$ TeV and sufficient integrated luminosity, $30-100 fb^{-1}$,  
through the production subprocess $q \bar q'\rightarrow WH$, followed by 
$W \rightarrow \ell\bar \nu$ and $H\rightarrow b\bar b$.  
Since the expected number of events is small, it is important to calculate 
the cross section as accurately as possible. In Ref.\cite{a05}  
the $ O(\alpha _s)$ QCD corrction 
to the total cross section to this process have been calculated, and the 
QCD correction were found to be about $12\%$ in the $\bar{MS}$ scheme  
at the Fermilab Tevatron and the LHC in the SM. In general, the SM electroweak corrections are small  
comparing with the QCD correction. Beyond the SM, the electroweak corrections  
might be enhanced, since more Higgs bosons with stronger couplings to top or  
bottom quarks are involved in some new physics models; for example, the 
minimal supersymmtric model(MSSM)\cite{a06} \cite{a11},  
which predict that the lightest Higgs 
boson $h_0$ be less than $140 GeV$. Therefore, it is worthwhile to calculate 
the electroweak corrections to the light Higgs boson production via 
$q\bar q'\rightarrow Wh_0$. In this paper we present the calculation of 
the top quark loop correction of order $\alpha m^2_t/m^2_W$ to 
the Higgs boson production at the Fermilab Tevatron in both the SM and 
the MSSM. These corrections arise from the virtual effects of the third 
family (top and bottom) of quark, neutral and charged Higgs bosons.  
And we shall present the complete calculations of the electroweak  
radiative corrections to this process in a future publication\cite{a07}.

\section{Caculations} 
The Feynman diagrams for the lightest Higgs boson production 
via $q(p_1) \bar q'(p_2)\rightarrow W (k_1) h_0(k_2)$ , which include 
the top quark loop corrections of order $\alpha m^2_t/m^2_W$ to the process 
$q\bar q'\rightarrow Wh_0$, are shown in Fig 1. We perform the calculation 
in the 't Hooft-Feynman gauge and use dimensional regularization to all 
the ultraviolet divergences in the virtual loop corrections utilizing  
the on-mass-shell renormalization\cite{a08}, in which the fine-structure constant 
$\alpha$ and the physical masses are chosen to be the renormalized  
paramenters, and the finite parts of the countertems are fixed by the  
renormalization conditions. As far as the parameters $\beta$ and $\alpha$, 
for the MSSM we are considering, they have to be renormalized, too. In the 
MSSM they are not independent. Neverthless, we follow the approach of Mendez 
and Pomarol\cite{a09} in which they consider them as independent renormalized  
parameters and fixed the correspoding renormalization constant by a  
renormalization condition that the on-mass-shell $H^+\bar \ell\nu _{\ell}$ 
and $h_0\bar \ell\ell$ couplings keep the forms of Eq.(3) of Ref.\cite{a09} to all order 
of perturbation theory. 
 
  We define the Mandelstam variables as 
\begin{eqnarray} 
\hat{s}&=&(p_1+p_2)^2=(k_1+k_2)^2\nonumber \\ 
\hat{t}&=&(p_1-k_1)^2=(p_2-k_2)^2 \nonumber \\ 
\hat{u}&=&(p_1-k_2)^2=(p_2-k_1)^2. 
\end{eqnarray} 
  
 The relevant renormalization constants are defined as 
\begin{eqnarray} 
m^2_{W0}=m^2_W+ \delta m^2_W, \ \ \ \  m^2_{Z0}=m^2_Z+ \delta m^2_Z,  
\end{eqnarray}  
\begin{eqnarray} 
\tan \beta_0=(1+\delta Z_{\beta})\tan \beta,\ \ \ \ 
\sin \alpha_0=(1+\delta Z_{\alpha})\sin \alpha, 
\end{eqnarray}  
\begin{eqnarray} 
W^{\pm \mu}_0=Z^{1/2}_WW^{\pm\mu}+ 
i Z^{1/2}_{H^\pm W}\partial^{\mu}H^{\pm},\ \ \ \ 
H^{\pm}_0=(1+\delta Z_{H^{\pm}})^{1/2}H^{\pm},  
\end{eqnarray}  
\begin{eqnarray} 
h_0=(1+\delta h_0)^{1/2}h+ 
Z^{1/2}_{h_0H}H,\ \ \ \ 
H_0=(1+\delta Z_H)^{1/2}H+Z^{1/2}_{Hh_0}h 
\end{eqnarray}

 Taking into account the $O(\alpha m_t^2/m_W^2)$ corrections, the renormalized  
amplitude for $q\bar q'\rightarrow Wh_0$  can be witten as 
\begin{eqnarray} 
M_{ren}=M_0+\delta M^{self}+\delta M^{vertex}, 
\end{eqnarray} 
where $M_0$ is the amplitude at the tree level, $\delta M^{self}$  
and $\delta M^{vertex}$ represent the corrections arising 
from the self-energy and vertex  diagrams, respectively. $M_0$ is given by 
\begin{eqnarray} 
M_0= {e^2 m_W \sin(\beta-\alpha) 
\over \sqrt{2} (m_W^2-\hat{s}) \sin\theta_w^2} 
\bar{d}(p_2)\rlap/\epsilon P_Lu(p_1), 
\end{eqnarray} 
where $P_{L,R}\equiv(1\mp\gamma_5)/2$. 
$\delta M^{self}$ is given by 
\begin{eqnarray} 
\delta M^{self} & = & {\delta m_W^2 +(m_W^2 -\hat{s}) \delta Z_W  
\over \hat{s} - m_W^2} M_0 \nonumber \\ 
&+&{ N_c e^4 m_{W} \sin(\beta-\alpha) \over 288 \sqrt{2} \pi^2 \hat{s} 
     ( - m_{W}^2 + \hat{s} )^2  \sin\theta_w^4} 
\left[ 6 \hat{s} m_{t}^2 - 2 \hat{s}^2 +  
       3 m_{t}^2 ( m_{t}^2 - 2 \hat{s} )\right. \nonumber \\  
      & \times & B_0(0,  m_b^2,  m_{t}^2)  
       +  \left. 
       3 ( - m_{t}^4 -\hat{s} m_{t}^2 +2 \hat{s}^2 ) 
        B_0(\hat{s},  m_b^2,  m_{t}^2) \right] \bar{d}(p_2)\rlap/\epsilon P_Lu(p_1) 
\end{eqnarray} 
with 
\begin{eqnarray} 
\delta m_W^2 &=& 
{N_c e^2 m_{t}^2 \over 96 \pi^2 \sin\theta_w^2} 
\left[ 
-2 + 2 B_0(0,m_b^2, m_{t}^2)-B_0( m_{W}^2,  m_b^2,m_{t}^2)-
4 B_0(0,m_t^2,m_t^2)
   \right. \nonumber\\ 
     &+& \left. 
{ m_{t}^2\over m_{W}^2} 
\left[B_0(0, m_b^2, m_{t}^2)-B_0( m_{W}^2, m_b^2, m_{t}^2)\right]  
\right],  
\end{eqnarray} 
\begin{eqnarray} 
\delta Z_{W} &=& 
{N_c e^2 \over 288 m_W^4  \pi^2  \sin\theta_w^2} 
\left[ 2 m_W^4+ 3 m_t^4 B_0(0, m_b^2, m_t^2) 
-3 (m_t^4+2 m_W^4) \right. \nonumber \\  
&\times &B_0(m_W^2, m_b^2, m_t^2) 
+\left. 
3 m_W^2 (m_t^4 + m_t^2 m_W^2-2 m_W^4) G(m_W^2, m_b^2, m_t^2)\right], 
\end{eqnarray} 
Here and below, $B_0$, $C_0$, $C_i$ and $C_{ij}$ 
is the two-point and three-point scalar integrals,  definitions for  
which can be found in 
Ref. \cite{denner} and $G$ is the derivative of $B_0$ which is expressed as 
\begin{eqnarray} 
G(M^2,M_1^2,M_2^2)={\partial B_0(k^2,M_1^2,M_2^2)\over 
 \partial k^2}|_{k^2=M^2}. 
\end{eqnarray} 
$\delta M^{vertex}$ is given by 
\begin{eqnarray} 
\delta M^{vertex}&=& M_0 \left[ {1\over 2} \delta Z_{h_0}+{\delta m_W^2 -\delta m_Z^2 
\over 2 (m_Z^2-m_W^2)} +{\delta m_Z^2 \over m_Z^2}+{\delta m_W^2 \over m_W^2} 
\right. 
\nonumber\\ 
&+& 
\left. 
\cot (\beta-\alpha)(Z_{Hh_0}^{1/2}+\sin\beta \cos\beta \delta Z_\beta - 
\tan\alpha \delta Z_\alpha)\right] \nonumber \\ 
&+&f_1^{vertex} \bar{d}(p_2)\rlap/\epsilon P_L u(p_1)\nonumber \\ 
&+&f_2^{vertex} \bar{d}(p_2)\rlap/p_1 P_L u(p_1) \epsilon.p_1\nonumber \\ 
&+&f_3^{vertex} \bar{d}(p_2)\rlap/p_1 P_L u(p_1) \epsilon.p_2, 
\end{eqnarray} 
with 
\begin{eqnarray} 
    \delta m_Z^2 &=& {N_c e^2 m_t^2 \over 3 \cos\theta_w^2 \pi^2} 
    \left[ 
      {B_0(0,m_{t}^2,m_{t}^2)\over  
    6 } -  
  { B_0(0, m_{t}^2,  m_{t}^2)\over  
    16 \sin\theta_w^2} -  
  {2 \sin\theta_w^2  B_0(0, m_{t}^2, m_{t}^2)\over  
    9}  
    \right. 
    \nonumber\\ 
    &-&  
    \left. 
  {  B_0( m_{Z}^2,m_{t}^2,  m_{t}^2)\over  
    6} -  
  {  B_0( m_{Z}^2, m_{t}^2, m_{t}^2)\over  
    32 \sin\theta_w^2} +  
  {2  \sin\theta_w^2 B_0(m_{Z}^2, m_{t}^2, m_{t}^2)\over  
    9} 
    \right], 
\end{eqnarray} 
\begin{eqnarray} 
\delta Z_{h_0} &=& 
{-N_c  e^2 \cos^2 (\alpha) \csc^2 (\beta) m_{t}^2 \over  
32  m_{W}^2 \pi^2  \sin\theta_w^2} 
     \left[  
     B_0(m_{h_0}^2, m_{t}^2, 
         m_{t}^2) \right. \nonumber \\ 
         &+&\left. (m_{h_0}^2-4 m_{t}^2) G( m_{h_0}^2, m_{t}^2,  m_{t}^2)  
          \right],  
\end{eqnarray} 
\begin{eqnarray} 
Z_{Hh_0}^{1/2}&=& {N_c e^2 m_t^2 \cos (\alpha) \csc^2(\beta) \sin (\alpha)\over  
   {32\,\left( {{{\it m_{H}}}^2} - {{{\it m_{h_0}}}^2} \right) \,{{{\it m_W}}^2}\, 
     {{\pi }^2}\,{{{\it \sin\theta_w}}^2}}} 
\left[ -2 m_t^2  
-  
       2 m_t^2 B_0(0, m_t^2, m_t^2)  
       \right. 
       \nonumber\\ 
       &+&  
       \left. 
        (m_{h_0}^2 -4 m_t^2) 
         B_0( m_{h_0}^2, m_t^2, m_t^2) \right],  
\end{eqnarray} 
\begin{eqnarray} 
\delta Z_\beta &=& {\delta m_Z^2 -\delta m_W^2 
\over 2 (m_Z^2-m_W^2)}  
-{\delta m_Z^2 \over 2 m_Z^2}+{\delta m_W^2 \over 2 m_W^2}- 
{1\over 2} \delta Z_{H^\pm} -{m_W\over \tan\beta} Z_{WH^\pm}^{1/2}, 
\end{eqnarray} 
\begin{eqnarray} 
\delta Z_\alpha&=& -\sin^2\beta \delta Z_\beta +{\delta m_Z^2 -\delta m_W^2 
\over 2 (m_Z^2-m_W^2)}  
-{\delta m_Z^2 \over 2 m_Z^2}+{\delta m_W^2 \over 2 m_W^2}- 
{1\over 2}\delta Z_{h_0} +{\cos\alpha\over \sin\alpha} Z_{Hh_0}^{1/2}, 
\end{eqnarray} 
\begin{eqnarray} 
\delta Z_{H^\pm} &=& 
{N_c g^2  m_{t}^2 \over 32 m_{W}^2 \pi^2 } 
\left[ 
 - B_0( m_{H^\pm}^2,  m_b^2, m_{t}^2)\cot^2(\beta)\right.\nonumber \\   
 &+&\left. 
       ( m_{t}^2 -m_{H^\pm}^2) \cot^2(\beta)  
       G( m_{H^\pm}^2,  m_b^2, m_{t}^2) \right], 
\end{eqnarray} 
\begin{eqnarray} 
Z_{H^\pm W}^{1/2}&=&  
{N_c  g^2 m_{t}^2 \cot (\beta) \over 32 m_{H^\pm}^2 m_{W}^3 \pi^2}  
\left[  m_{t}^2 
B_0(0, m_b^2,  m_{t}^2) \right. \nonumber \\ 
&+&\left.  
         (m_{H^\pm}^2-m_{t}^2) 
        B_0(  m_{H^\pm}^2, m_b^2,  m_{t}^2) 
        \right], 
\end{eqnarray} 
\begin{eqnarray} 
f_1^{vertex}&=&{-N_c e^4  m_{t}^2 \over {32\,{\sqrt{2}}\,{  m_{W}}\,{{\pi }^2}\,\left( {{{  m_{W}}}^2} - \hat{s} \right) \, 
     {{{  \sin\theta_w}}^4}}} 
     \left[ 
      -2   B_0(\hat{s}, m_b^2,  m_{t}^2)  
      \right. 
      \nonumber\\ 
      &+&  
      ( -2  m_{t}^2 - \hat{t} )  C_0( 
        m_{h_0}^2,  m_{W}^2,\hat{s}, m_{t}^2, 
          m_{t}^2,  m_b^2) \nonumber \\ 
          &+&  
         ( -2 m_{W}^2 - \hat{t} )  C_1(  m_{W}^2,\hat{s}, m_{h_0}^2 , 
          m_{t}^2, m_b^2,  m_{t}^2 ) \nonumber\\ 
          &+&  
         ( - m_{h_0}^2 - 2 \hat{t} ) C_2( m_{W}^2,\hat{s},  m_{h_0}^2, 
         m_{t}^2,  m_b^2,  m_{t}^2 ) \nonumber\\ 
         &+&  
       \left. 
       4 C_{00}(m_{W}^2,\hat{s},  m_{h_0}^2 , 
           m_{t}^2,  m_b^2,  m_{t}^2 ) \right],  
      \end{eqnarray} 
     \begin{eqnarray} 
 f_2^{vertex}&=& 
{-N_c \ e^4  m_{t}^2 \over  
{16\,{\sqrt{2}}\,{  m_{W}}\,{{\pi }^2}\,\left( {{{  m_{W}}}^2} - \hat{s} \right) \, 
     {{{  \sin\theta_w}}^4}}} 
  \left[ 
   -  C_0( m_{h_0}^2,m_{W}^2,\hat{s}, 
          m_{t}^2,  m_{t}^2,  m_b^2)  
          \right. 
          \nonumber\\ 
          &-&  
        C_1( m_{W}^2,\hat{s}, m_{h_0}^2, 
          m_{t}^2, m_b^2, m_{t}^2 ) \nonumber\\ 
          & -&  
       3 C_2( m_{W}^2,\hat{s},  m_{h_0}^2 , 
          m_{t}^2,  m_b^2,  m_{t}^2 ) \nonumber \\ 
          &- & 
       2  C_{12}( m_{W}^2,\hat{s},  m_{h_0}^2, 
           m_{t}^2,  m_b^2, m_{t}^2 ) \nonumber \\ 
           &-&  
           \left. 
       2 C_{22}(m_{W}^2,\hat{s},  m_{h_0}^2, 
         m_{t}^2,  m_b^2,  m_{t}^2 ) \right], 
     \end{eqnarray} 
     \begin{eqnarray} 
f_3^{vertex}&=& 
{-N_c  e^4  m_{t}^2 \over  
   {16\,{\sqrt{2}}\,{  m_{W}}\,{{\pi }^2}\,\left( {{{  m_{W}}}^2} - \hat{s} \right) \, 
     {{{  \sin\theta_w}}^4}}} 
\left[ 
 - C_2(m_{W}^2, \hat{s},  m_{h_0}^2, 
         m_{t}^2,  m_b^2,  m_{t}^2 ) \right. \nonumber \\ 
         &-&  
       2  C_{12}( m_{W}^2,\hat{s},  m_{h_0}^2, 
          m_{t}^2, m_b^2, m_{t}^2 ) \nonumber \\ 
          &-&  
          \left. 
       2  C_{22}( m_{W}^2,\hat{s},  m_{h_0}^2 , 
          m_{t}^2,  m_b^2,  m_{t}^2 ) \right]. 
 \end{eqnarray} 
  
 The corresponding amplitude squared for the process  
 $q \bar q'   
\rightarrow W h_0 $ 
 can be written as 
\begin{eqnarray} 
\bar{{\sum}}\left| M_{ren}\right|^2=\bar{{\sum}}\left| M_{0}\right|^2 
 +2 Re \bar{{\sum}}(\delta M^{self}+ \delta M^{vertex}) M_0^\dagger, 
\end{eqnarray} 
where the bar over the summation recalls average over initial partons spins. 
The cross section of  process $q \bar q'   
\rightarrow W h_0 $ is 
\begin{eqnarray} 
\hat{\sigma} =\int_{\hat{t}_{min}}^{\hat{t}_{max} }{1\over 16\pi \hat{s }^2}\bar 
{ \sum_{ spins} }\left|  M\right|^2 d \hat{t} 
\end{eqnarray} 
with 
\begin{eqnarray} 
\hat{t}_{min}&=&{m_{h_0}^2+m_W^2-\hat{s}\over 2}-\sqrt{ 
   (\hat{s}-(m_{h_0}+m_W)^2)(\hat{s}-(m_{h_0}-m_W)^2)/2} \nonumber \\ 
\hat{t}_{max}&=&{m_{h_0}^2+m_W^2-\hat{s}\over 2}+\sqrt{ 
   (\hat{s}-(m_{h_0}+m_W)^2)(\hat{s}-(m_{h_0}-m_W)^2)/2}. 
\end{eqnarray} 
The total cross section of $P\bar{P}\rightarrow q \bar q' \rightarrow  
W h_0$ 
can be obtained by folding the $\hat{\sigma}$  with the 
parton luminosity 
\begin{eqnarray} 
\sigma (s)=\int^{1}_{(m_{h_0}+m_W)/\sqrt{s}} dz 
{dL\over dz} \hat{\sigma} (q \bar q'\rightarrow W h_0 \mbox{ at 
$\hat{s}=z^2 s$}), 
\end{eqnarray} 
where $\sqrt{s}$ and $\sqrt{\hat{s}}$ is the CM energy of $P\bar{P}$ and 
$q \bar q'$, respectively, and $dL/dz$ is the paron 
luminosity, which is defined as 
\begin{eqnarray} 
{dL\over dz}=2 z \int_{z^2}^1{dx\over x} 
f_{q /P}(x,q^2) f_{q' /\bar{P}}(z^2/x,q^2), 
\end{eqnarray} 
where $f_{q/P}(x,q^2)$ and $f_{q'/\bar{P}}(z^2/x,q^2)$ are  
the parton distribution function \cite{CTEQ}.

\section{Numerical Results} 
  In the following we present some numerical results.  
In our numerical calculations, the SM parameters were taken to be  
$m_W=80.33 GeV$, $m_Z=91.187 GeV$, $m_t=176 GeV$, $m_b=4.5 GeV$ and  
$\alpha(m_W)={1\over 128}$. Moreover, we use the relation\cite{a11} between  
the Higgs boson masses $m_{h_0,H,A,H^{\pm}}$ and parameters  
$\alpha, \beta$ at one-loop, and choose $m_{h_0}$ and $\tan \beta$ as 
two independent input parameters. As explained in Ref.\cite{a09}, there is a small 
inconsistency in doing so since the parameters $\alpha$ and $\beta$ of  
Ref.\cite{a11} are not the ones defined by the conditions given by Eq.(3)  
of Ref.\cite{a09}. Nevertheless, this difference would only induce a higher order 
change\cite{a09}. We will limit the value of $\tan \beta$ to be 
in the ranges $2\leq \tan \beta\leq 30$, which are consistent with ones  
required by the most popular MSSM model with scenarios motivated by current  
low energy data (including $\alpha _s$, $R_b$ and the branching ratio of  
$b\rightarrow s\gamma$). 
    
  In Fig.2 we present the tree-level 
total cross sections versus the Higgs boson mass in both the SM and the 
MSSM for the different values of $\tan \beta$, using the CTEQ3L parton 
distributions\cite{CTEQ}. Figure 2 shows the total cross sections in the SM always 
are larger than ones in the MSSM, and they are almost same only for  
$\tan \beta=2$, or the mass of the Higgs boson approch to $130$ GeV. 
   
  In Fig.3 we shows the top quark loop corrections of order  
$\alpha m_t^2/m_W^2$ to the total cross sections. From Fig.3 one sees that 
in the SM the corrections are not sensitive to the mass of the Higgs 
boson and amounts to $1\% \sim 2\%$ reduction in the cross section. And 
in the MSSM the corrections depend strongly on the values of $m_{h_0}$ for 
all $\tan \beta$. Especially for $\tan \beta >2$, such correction can reach  
about $-4\%$ when $m_{h_0}=60 GeV$, but the correction is only about $-1\%$ 
if $m_{h_0}=130 GeV$. Since QCD correction increases the tree-level total  
cross sections by about $12\%$\cite{a05}, it is necessary  
for an accurate calculation of  
the cross sections   
to include the top 
quark loop corrections, which typically imply a few percent reduction in 
the cross sections. 
 
  In conclusion, we have calculated that the top quark loop corrections of  
order $\alpha m_t^2/m_W^2$ to the neutral Higgs boson production via  
$q \bar q' \rightarrow WH$ at the Fermilab Tevatron in the SM and the MSSM.  
In contrast to the QCD corrections, such corrections reduce the tree-level  
total cross sections by about $1\% \sim 2\%$ in the SM,  
and $1\% \sim 4\%$ in the MSSM.

\section*{Acknowledgements} 
This work was supported in part by the National Natural Science Foundation 
of China and a grant from the State Commission of Science and  
Technology of China.

\input{fig}

\end{document}

%% file: fig.tex
\newpage
\begin{figure}
\epsfxsize=12 cm
\centerline{\epsffile{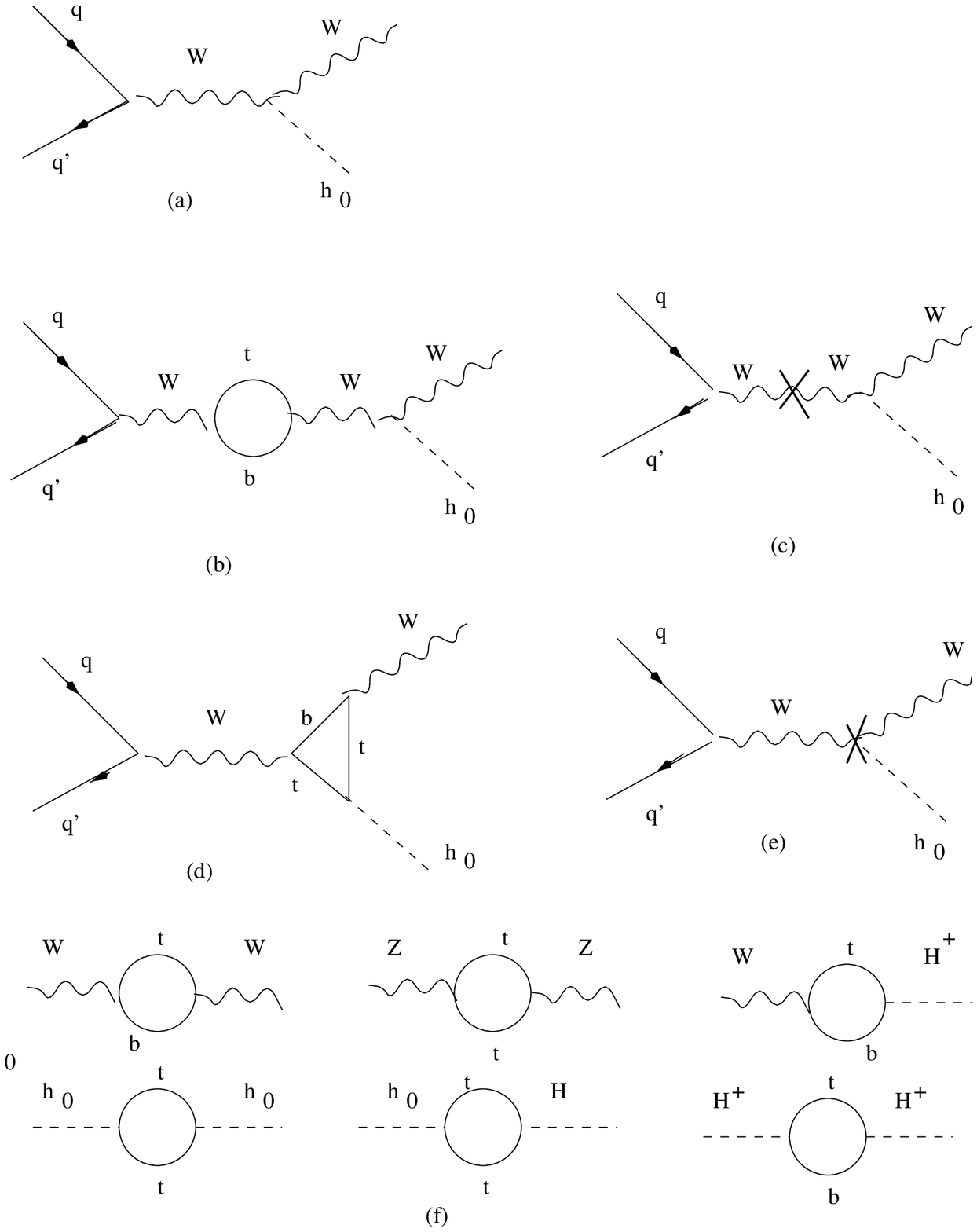}}
\caption[]{Feynmann diagrams for the
 process $q \bar q' \rightarrow W h_0$.
}
\end{figure}

\begin{figure}
\epsfxsize=12 cm
\centerline{\epsffile{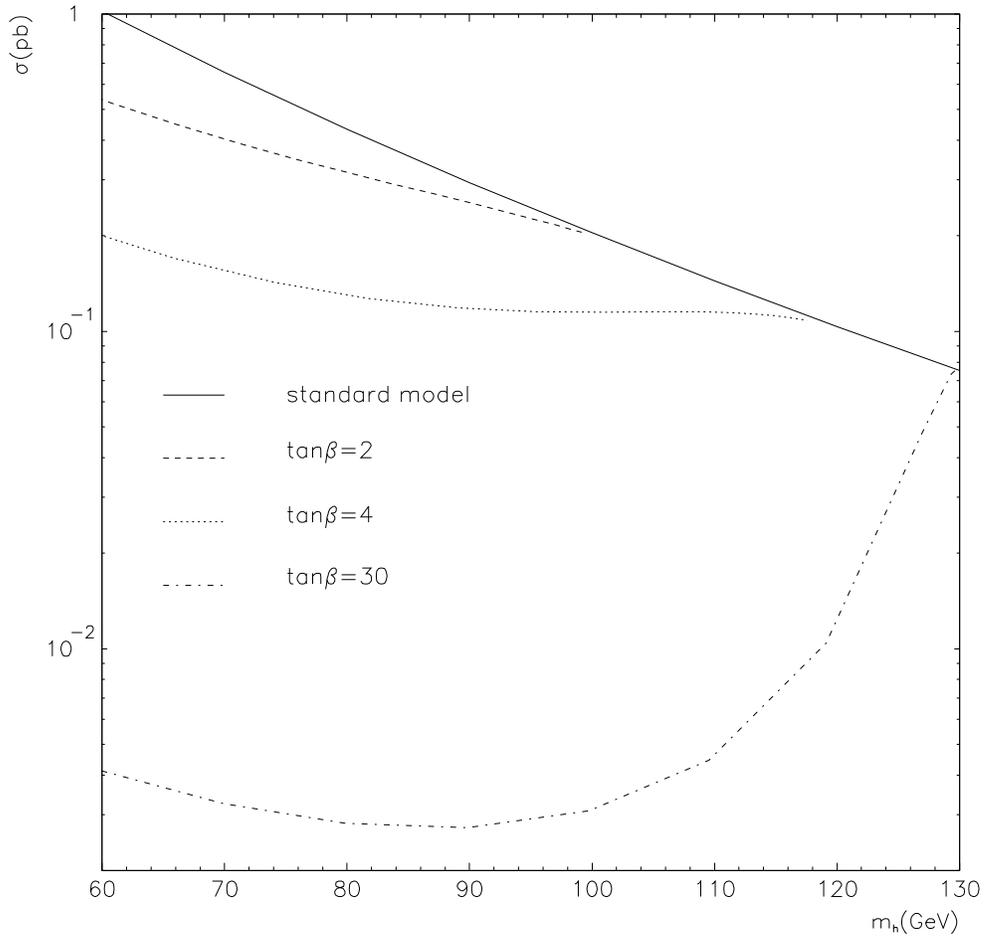}}
\caption[]{Tree-level cross sections as a function of the Higgs boson
mass of the process $q \bar q' \rightarrow W h_0$ 
with $\sqrt{s}=2 TeV$ at Tevatron.}
\end{figure}

\begin{figure}
\epsfxsize=12 cm
\centerline{\epsffile{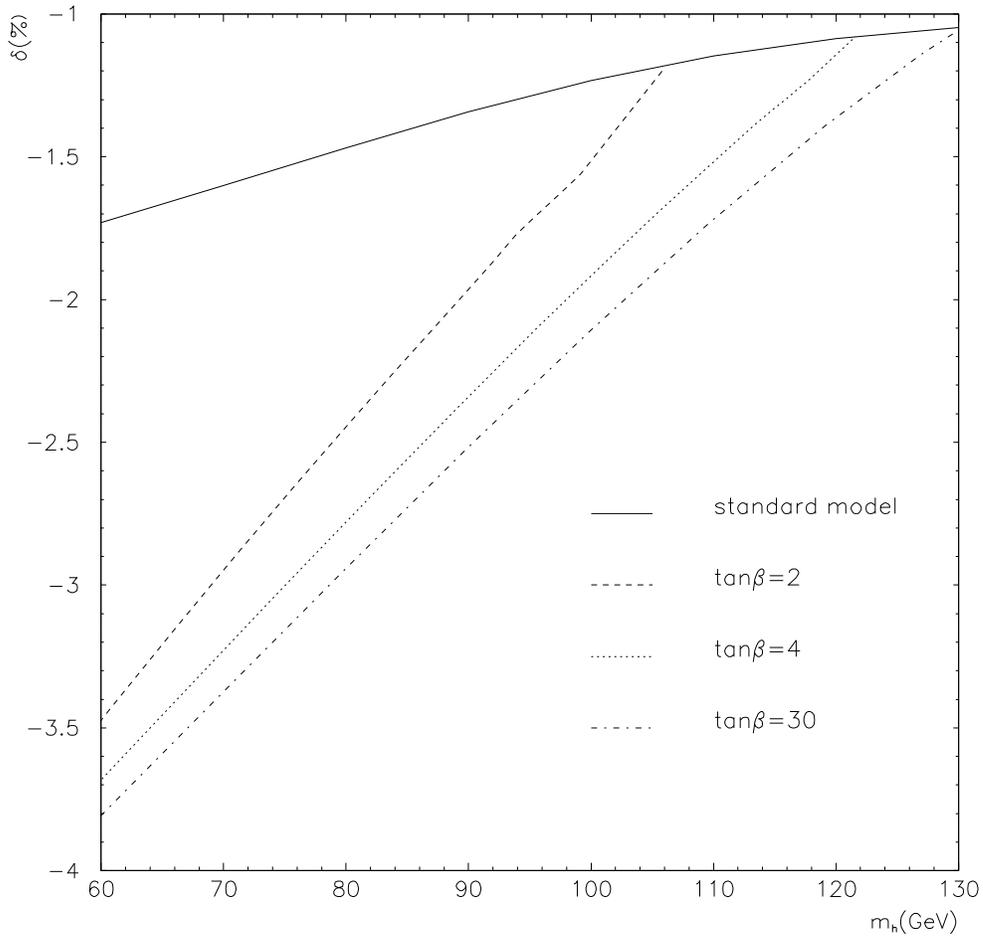}}
\caption[]{Relative corrections $\delta \sigma /\sigma_0 $ as a function of the Higgs boson
mass of the process $q \bar q' \rightarrow W h_0$ 
with $\sqrt{s}=2 TeV$ at Tevatron.}
\end{figure}